# Dynamic Density Response of Trapped Interacting Quantum Gases[*]

Renu Bala[1], J. Bosse[2], K.N. Pathak[1]

[1]*Department of Physics, Panjab University, Chandigarh, India*
[2]*Institute for Theoretical Physics, Freie Universität Berlin, Germany*

**Abstract.** An expression for the dynamic density response function has been obtained for an interacting quantum gas in Random Phase Approximation (RPA) including first order self and exchange contribution. It involves the single particle wave functions and eigen values. The expression simplifies when diagonal elements are considered. The diagonal elements of the imaginary part of Fourier transformed response function is relevant in the measurement of Bragg scattering cross-section and in several other applications.

## INTRODUCTION

Trapped quantum gases have interesting properties due to finite number of particles, its temperature, and inter-particle interactions. Various aspects of these have been studied during the last fifteen year [1, 2, 3, 4]. In this paper we present the derivation of the density response function of inhomogeneous quantum gases using the Green's function method treating the interaction in the Random Phase Approximation (RPA) including first order self energy and exchange contribution to it.

## THEORETICAL FORMALISM

We consider a system of N particles. The Hamiltonian for the system in the second quantized form is given as $H = H_0 + H_1$,

$$H_0 = \sum_{\lambda\alpha}(\varepsilon_{\lambda\alpha} - \mu)a^\dagger_{\lambda\alpha}a_{\lambda\alpha}$$

$$H_1 = \frac{1}{2}\sum_{\substack{\lambda,\lambda',\mu',\mu \\ \alpha,\alpha'}} W^{\mu',\mu}_{\lambda\lambda'} a^\dagger_{\lambda\alpha} a^\dagger_{\lambda'\alpha'} a_{\mu'\alpha'} a_{\mu\alpha}, \quad (1)$$

where $W^{\mu',\mu}_{\lambda,\lambda'} = \int d\vec{r}d\vec{r}'\varphi^*_\lambda(\vec{r})\varphi^*_{\lambda'}(\vec{r}')V(|\vec{r}-\vec{r}'|)\varphi_{\mu'}(\vec{r}')\varphi_\mu(\vec{r})$ is the matrix element of inter-atomic potential $V(|\vec{r}-\vec{r}'|)$ (taken as central) between the single particles states defined by $\varphi_\lambda$'s corresponding to eigen values $\varepsilon_\lambda^s$. In Eq. (1) α's are the spin components corresponding to particle of spin s. Henceforth we suppress the explicit consideration of spin degrees of freedom.

We define the retarded response function as [5]

$$\bar{\chi}(\vec{r},\vec{r}',t) = -i\theta(t)\langle[\rho(\vec{r},t),\rho^\dagger(\vec{r}',0)]\rangle,$$
$$= \sum_{\nu',\nu}\langle\langle\rho_{\nu',\nu}(\vec{r},t);\rho^\dagger(\vec{r}',0)\rangle\rangle = \sum_{\nu',\nu}\bar{G}_{\nu',\nu}(\vec{r},\vec{r}',t), \quad (2)$$

where $\bar{G}$ is the Green's function. In Eq.(2) $\rho_{\nu',\nu}(\vec{r},t) = \varphi^*_{\nu'}(\vec{r})\varphi_\nu(\vec{r})a^\dagger_{\nu'}(t)a_\nu(t)$ is the partial density operator which on summing over ν', ν gives the density operator $\rho(\vec{r})$, $\theta(t)$ is the unit-step function and the angular brackets denote the grand canonical ensemble average. The Green's function $\bar{G}_{\nu',\nu}(\vec{r},\vec{r}',t)$ satisfies the equation of motion,

$$i\frac{d\bar{G}_{\nu',\nu}(\vec{r},\vec{r}',t)}{dt} = \delta(t)<[\rho_{\nu',\nu}(\vec{r},t),\rho^\dagger(\vec{r}',0)]> +$$
$$\langle\langle[\rho_{\nu',\nu}(\vec{r},t),H_0(t)];\rho^\dagger(\vec{r}',0)\rangle\rangle + \langle\langle[\rho_{\nu',\nu}(\vec{r},t),H_1(t)];\rho^\dagger(\vec{r}',0)\rangle\rangle. \quad (3)$$

Evaluation of the commutators in the first and second term is straight forward. The last commutator in Eq. (3) contains four terms, which on using the fact that the matrix elements of inter atomic potential is symmetric under particle interchange, reduces to

$$\varphi^*_{\nu'}(\vec{r})\varphi_\nu(\vec{r})\sum_{\lambda',\mu',\mu}W^{\mu',\mu}_{\nu,\lambda'}a^\dagger_{\nu'}a^\dagger_{\lambda'}a_{\mu'}a_\mu$$
$$-\varphi^*_{\nu'}(\vec{r})\varphi_\nu(\vec{r})\sum_{\lambda',\mu',\mu}W^{\mu',\nu}_{\mu,\lambda'}a^\dagger_\mu a^\dagger_{\lambda'}a_{\mu'}a_\nu \quad (4)$$

Substituting (4) in Eq. (3) gives higher order Green's function. Equation of motion for these Green's functions will lead to still higher order Green's function and so on. In order to close the infinite hierarchy of coupled equations, we truncate this Green's function using the approximation,

$$\langle\langle a^\dagger_1 a^\dagger_2 a_3 a_4;\rho^\dagger\rangle\rangle = \langle a^\dagger_1 a_4\rangle\langle\langle a^\dagger_2 a_3;\rho^\dagger\rangle\rangle + \langle a^\dagger_2 a_3\rangle\langle\langle a^\dagger_1 a_4;\rho^\dagger\rangle\rangle$$
$$+ \eta\langle a^\dagger_1 a_3\rangle\langle\langle a^\dagger_2 a_4;\rho^\dagger\rangle\rangle + \eta\langle a^\dagger_2 a_4\rangle\langle\langle a^\dagger_1 a_3;\rho^\dagger\rangle\rangle, \quad (5)$$

where η=±1 for bosons & fermions respectively.

In Eq. (5) first & second terms gives RPA & Hartree (dynamical) contributions respectively, third & fourth gives the first order self & exchange contribution to RPA. Using the approximation Eq. (5) in Eq. (3) & taking the Fourier transform w. r. t time thereafter summing over $\nu'$, $\nu$ we get,

$$\chi(\vec{r},\vec{r}',\omega) = \chi_0(\vec{r},\vec{r}',\omega) + \int d\vec{r}'' d\vec{r}''' \chi_0(\vec{r},\vec{r}'',\omega) V(|\vec{r}''-\vec{r}'''|) \chi(\vec{r}''',\vec{r}',\omega) +$$

$$\sum_{\nu,\nu',\mu',\mu} \int d\vec{r}'' d\vec{r}''' \{ P_{\mu',\nu,\nu'}(\vec{r},\vec{r}'',\vec{r}''',\omega) + \eta Q_{\mu',\nu,\nu'}(\vec{r},\vec{r}'',\vec{r}''',\omega) + \eta R_{\mu',\nu,\nu'}(\vec{r},\vec{r}'',\vec{r}''',\omega)$$

$$- S_{\nu,\mu',\mu}(\vec{r},\vec{r}'',\vec{r}''',\omega) - \eta T_{\nu,\mu',\mu}(\vec{r},\vec{r}'',\vec{r}''',\omega) - \eta U_{\nu,\mu',\mu}(\vec{r},\vec{r}'',\vec{r}''',\omega) \} G_{\nu',\mu}(\vec{r}''',\vec{r}',\omega) \quad (6),$$

where $\omega$ stands for $\omega + \iota 0$ and $\bar{n}_\nu$ is the average occupation number for the state $\nu$, symbols are defined

$$\chi_0(\vec{r},\vec{r}',\omega) = \sum_{\nu,\nu'} \frac{\varphi_\nu^*(\vec{r})\varphi_{\nu'}(\vec{r})\varphi_{\nu'}^*(\vec{r}')\varphi_\nu(\vec{r}')(\bar{n}_{\nu'} - \bar{n}_\nu)}{(\omega - \varepsilon_\nu + \varepsilon_{\nu'})},$$

$$P_{\mu',\nu,\nu'}(\vec{r},\vec{r}'',\vec{r}''',\omega) = \frac{\varphi_\nu^*(\vec{r})\varphi_\nu(\vec{r})\varphi_{\nu'}^*(\vec{r}'')\varphi_{\mu'}^*(\vec{r}''')\varphi_{\mu'}(\vec{r}''')V(|\vec{r}''-\vec{r}'''|)\bar{n}_{\mu'}}{\varphi_{\nu'}^*(\vec{r}'')(\omega - \varepsilon_\nu + \varepsilon_{\nu'})},$$

$$Q_{\mu',\nu,\nu'}(\vec{r},\vec{r}'',\vec{r}''',\omega) = \frac{\varphi_\nu^*(\vec{r})\varphi_\nu(\vec{r})\varphi_{\nu'}^*(\vec{r}'')\varphi_{\nu'}^*(\vec{r}''')\varphi_{\mu'}(\vec{r}''')V(|\vec{r}''-\vec{r}'''|)\bar{n}_{\mu'}}{\varphi_{\nu'}^*(\vec{r}'')(\omega - \varepsilon_\nu + \varepsilon_{\mu'})},$$

$$R_{\mu',\nu,\nu'}(\vec{r},\vec{r}'',\vec{r}''',\omega) = \frac{\varphi_{\nu'}^*(\vec{r})\varphi_\nu(\vec{r})\varphi_\nu^*(\vec{r}''')\varphi_{\mu'}^*(\vec{r}''')\varphi_{\mu'}(\vec{r}''')V(|\vec{r}''-\vec{r}'''|)\bar{n}_{\mu'}}{\varphi_{\nu'}^*(\vec{r}'')(\omega - \varepsilon_\nu + \varepsilon_{\nu'})},$$

$$S_{\nu,\mu',\mu}(\vec{r},\vec{r}'',\vec{r}''',\omega) = \frac{\varphi_\nu^*(\vec{r})\varphi_\mu(\vec{r})\varphi_{\mu'}^*(\vec{r}'')\varphi_{\mu'}(\vec{r}''')\varphi_\nu(\vec{r}''')V(|\vec{r}''-\vec{r}'''|)\bar{n}_{\mu'}}{\varphi_\mu^*(\vec{r}'')(\omega - \varepsilon_\mu + \varepsilon_\nu)},$$

$$T_{\nu,\mu',\mu}(\vec{r},\vec{r}'',\vec{r}''',\omega) = \frac{\varphi_\nu^*(\vec{r})\varphi_\mu(\vec{r})\varphi_{\mu'}^*(\vec{r}'')\varphi_{\mu'}(\vec{r}''')\varphi_\nu(\vec{r}''')V(|\vec{r}''-\vec{r}'''|)\bar{n}_{\mu'}}{\varphi_\mu(\vec{r}''')(\omega - \varepsilon_\mu + \varepsilon_\nu)},$$

$$U_{\nu,\mu',\mu}(\vec{r},\vec{r}'',\vec{r}''',\omega) = \frac{\varphi_{\mu'}^*(\vec{r})\varphi_\nu(\vec{r})\varphi_\nu^*(\vec{r}'')\varphi_{\mu'}(\vec{r}''')\varphi_\mu(\vec{r}''')V(|\vec{r}''-\vec{r}'''|)\bar{n}_\nu}{\varphi_\mu(\vec{r}''')(\omega - \varepsilon_\nu + \varepsilon_{\mu'})}.$$

Now the integral equation obtained in Eq. (6) can be solved for the Green's function $G_{\nu',\mu}(\vec{r}'',\vec{r}',\omega)$ by iteration and we obtain,

$$\chi(\vec{r},\vec{r}',\omega) = \chi_0(\vec{r},\vec{r}',\omega) + \int d\vec{r}'' d\vec{r}''' \chi_0(\vec{r},\vec{r}'',\omega) V(|\vec{r}''-\vec{r}'''|) \chi(\vec{r}''',\vec{r}',\omega) +$$

$$\sum_{\nu,\nu',\mu,\mu'} \int d\vec{r}'' d\vec{r}''' \{ P_{\mu',\nu,\nu'}(\vec{r},\vec{r}'',\vec{r}''',\omega) + \eta Q_{\mu',\nu,\nu'}(\vec{r},\vec{r}'',\vec{r}''',\omega) + \eta R_{\mu',\nu,\nu'}(\vec{r},\vec{r}'',\vec{r}''',\omega)$$

$$- S_{\nu,\mu',\mu}(\vec{r},\vec{r}'',\vec{r}''',\omega) - \eta T_{\nu,\mu',\mu}(\vec{r},\vec{r}'',\vec{r}''',\omega) - \eta U_{\nu,\mu',\mu}(\vec{r},\vec{r}'',\vec{r}''',\omega) \}$$

$$\{ G_{0\nu',\mu}(\vec{r}''',\vec{r}',\omega) + \int d\vec{R}'' d\vec{R}''' G_{0\nu',\mu}(\vec{r}''',\vec{R}'',\omega) V(|\vec{R}''-\vec{R}'''|) \chi(\vec{R}''',\vec{r}',\omega) \} \quad (7),$$

Further we define

$$F_1(\vec{r},\vec{r}',\omega) = \sum_{\nu',\nu,\mu,\mu'} P_{\mu',\nu,\nu'} G_{0\nu',\mu}, \quad F_2(\vec{r},\vec{r}',\omega) = \sum_{\nu',\nu,\mu,\mu'} Q_{\mu',\nu,\nu'} G_{0\nu',\mu}$$

$$F_3(\vec{r},\vec{r}',\omega) = \sum_{\nu',\nu,\mu,\mu'} R_{\mu',\nu,\nu'} G_{0\nu',\mu}, \quad F_4(\vec{r},\vec{r}',\omega) = \sum_{\nu',\nu,\mu,\mu'} S_{\nu,\mu',\mu} G_{0\nu',\mu}$$

$$F_5(\vec{r},\vec{r}',\omega) = \sum_{\nu',\nu,\mu,\mu'} T_{\nu,\mu',\mu} G_{0\nu',\mu}, \quad F_6(\vec{r},\vec{r}',\omega) = \sum_{\nu',\nu,\mu,\mu'} U_{\nu,\mu',\mu} G_{0\nu',\mu}$$

With the above notations, Eq. (7) can be written in a matrix (for $\vec{r}$, $\vec{r}'$) form as,

$$\chi = (\chi_0 + F_1 + \eta F_2 + \eta F_3 - F_4 - \eta F_5 - \eta F_6)(1 - \chi_0 V - F_1 V - \eta F_2 V - \eta F_3 V + F_4 V + \eta F_5 V + \eta F_6 V)^{-1}$$

$$= (\chi_0 + \chi_1)[1 - V(\chi_0 + \chi_1)]^{-1}, \quad (8)$$

where $\chi_1 = (F_1 + \eta F_2 + \eta F_3 - F_4 - \eta F_5 - \eta F_6)$ is the first order contribution to RPA. This is believed to be new result [6] for Quantum gases and valid for any dimension. It provides dynamics beyond RPA where $\chi_1 = 0$. The interaction potential $V(|\vec{r}-\vec{r}'|)$ can be considered to be an effective two-body potential obtained from S-matrix theory or simply bare potential. It can be seen that for contact potential result simplifies.

## DISCUSSION

The expression for the density response function given in Eq. (8) is very general and is expected to be quite useful in discussing the dynamical properties of quantum gases under various situations at least for small interaction parameter. It contains dynamics beyond RPA. It is easy to obtain Eq. (8) in the momentum space and with the help of fluctuation, dissipation theorem; the diagonal element of the imaginary part of the response function can be used to obtain the Bragg scattering cross-section. Further it has been checked that for the homogenous interacting electron gas our approximation yields the result first obtained by Geldart and Taylor [7] and it has also been found that Eq. (8) gets simplified for homogenous interacting quantum atomic gases for any temperature. We note that we have already calculated the density response function for an inhomogeneous Bose gas in the high temperature limit [8]. The numerical result for interacting inhomogeneous quantum atomic gases will be presented elsewhere.

## ACKNOWLEDGMENT

Renu acknowledges CSIR, New Delhi for the grant of Junior Research Fellowship and K N Pathak thanks INSA for the award of senior scientist position.

## REFERENCES

* presented to ICACNM 23-26 Feb 2011 at Panjab University, Chandigarh, India.
1. C.J.Pethick and H.Smith, Bose-Einstein Condensation in Dilute Gases. Cambridge University Press, Cambridge, 2002.
2. Lev Pitaevskii and Sandro Stringari, Bose-Einstein Condensation, Clarendon Press, Oxford, 2003.
3. A.S. Parkins, D.F. Walls, Physics Reports **303**, 1 (1998).
4. M. Greiner, C.A. Regal, and J.T. Stewart, Phys. Rev.Lett, **94**, 110401, (2005)
5. D.N.Zubarev, Soviet Phy. Uspekhi, **3**, 3(1960).
6. Renu Bala, R. K.Moudgil, Sunita Srivastava & K.N.Pathak, Proceedings of 55th DAE Solid State Physics Symposium (2010) & PU M Phil. Thesis of Zile Singh Balhra(1982).
7. D. J. W. Geldart and R. Taylor, Can. J. Phys. **48**,155 (1970), ibid **48**, 167(1970).
8. Carol Jai Cruz, C. N. Kumar, K. N. Pathak, and J. Bosse, PRAMANA-Journal of Physics, **74**, 83 (2010).